\documentclass{article}
\usepackage{amsfonts}
\usepackage{amssymb}
\usepackage{amsmath}
\usepackage[dvips]{graphics}
\usepackage{latexsym}
\usepackage{subfigure}

\newtheorem{The}{Theorem}
\newtheorem{Lem}{Lemma}

\newcommand{\bv}{\begin{vmatrix}}
\newcommand{\ev}{\end{vmatrix}}
\newcommand{\pa}{\partial}

\newcommand{\bea}{\begin{eqnarray*}}
\newcommand{\eea}{\end{eqnarray*}}
\newcommand{\be}{\begin{eqnarray}}
\newcommand{\ee}{\end{eqnarray}}
\newcommand{\eqs}[1]{Eqs. (\ref{#1})}
\newcommand{\eq}[1]{Eq. (\ref{#1})}
\newcommand{\meq}[1]{(\ref{#1})}

\newcommand{\eqn}{&=&}

\newcommand{\oh}{\frac{1}{2}}

\title{Mass Dependence of the Entropy Product and Sum}
\author{Yuan Zhang\thanks{zhangyuan@mail.bnu.edu.cn}, Sijie Gao\thanks{Corresponding author. sijie@bnu.edu.cn}\\
Department of Physics, Beijing Normal University,\\
Beijing 100875, China}
\begin{document}
\maketitle

\begin{abstract}
For black holes with multiple horizons, the area product of all horizons has been proven to be mass independent in many cases. Counterexamples were also found  in some occasions. In this paper, we first prove a theorem derived from the first law of black hole thermodynamics and a mathematical lemma related to the Vandermonde determinant. With these arguments, we develop some general criteria for the mass independence of the  entropy product as well as the entropy sum. In particular, if a $d$-dimensional spacetime is spherically symmetric and  the radial metric function $f(r)$ is a Laurent series in $r$ with the lowest power  $-m$ and the highest power  $n$, we find the criterion is extremely simple: The entropy product  is mass independent if and only if $m\geq d-2$ and $n\geq4-d$. The entropy sum is mass independent if and only if $m\geq d-2$ and $n\geq 2$. Compared to previous works, our method does not require an exact expression of the metric.

Our arguments turn out to be useful even for rotating black holes. By applying our theorem and lemma to a Myers-Perry black hole with spacetime dimension $d$, we show that the entropy product/sum is mass independent for all $d>4$, while it is mass dependent only for $d=4$, i.e., the Kerr solution. \\

PACS number(s): 04.70.Dy, 04.50.Gh
\end{abstract}

\section{Introduction}
For an axisymmetric and stationary Einstein-Maxwell black hole with angular momentum $J$ and charge $Q$,  Marcus Ansorg and Jorg Hennig\cite{ansorg}\cite{ansorg2} proved the universal relation
\be
A^{+}A^{-}=(8\pi J)^{2}+(4\pi Q^{2})^{2},
\ee
 where $A^{+}$ and $A^{-}$ denote the areas of the event and Cauchy horizon, respectively.

Cveti\v{c} \emph{et al}. \cite{cvetic} generalized the calculation to higher-dimensional black holes with multihorizons. By explicit computation, they showed that the product of all horizon entropies  for rotating multicharge black holes in four
and higher dimensions is independent of the mass, in either asymptotically flat or asymptotically anti-de Sitter spacetimes. This subject has been further explored in recent years\cite{chen}-\cite{visser}. Recently, the entropy sum of all black hole horizons has been investigated\cite{meng-sum, yu}. In many known solutions, the sum is also mass independent.

So far, most studies on this issue rely on specific forms of metrics as well as explicit expressions of black hole mass, angular momentum, etc. In this paper, we aim to find some general criteria for the mass independence of the entropy product/sum. By employing the first law of black hole thermodynamics and the Vandermonde determinant, we find a very simple criterion for spherically symmetric black holes. We show that if the radial metric function $f(r)$ is a Laurent series then whether the entropy product/sum is mass independent is determined only by the lowest power and highest power of the series. Our arguments are also helpful for rotational black holes. We find that the entropy product for a Myers-Perry black hole is mass independent for all dimensions and the entropy sum is mass independent for all $d>4$, where $d$ is the spacetime dimension. Our calculation requires the expressions of entropy and temperature, but the explicit forms of the mass, angular momentum and charge are not needed.

\section{Entropy product, entropy sum and the first law}
In this section, we apply the first law to all horizons of the black hole and find some formulas related to the entropy product and entropy sum. A stationary black hole with mass $M$, charged $Q$ and angular momentum $J$ may have multiple horizons. Each horizon possesses a different temperature $T_{i}$, angular velocity $\Omega_i$, electrostatic potential $\Omega_i$ and entropy $S_{i}$.
The first law of black hole thermodynamics for each horizon reads
\be
dM=T_idS_i+\Omega_i dJ+\Phi_i dQ \,,
\ee
which obviously yields
\be
\frac{\partial S_{i}}{\partial M}=\frac{1}{T_{i}}\,.
\ee
In this paper, we require $T_i\neq 0$ for all horizons.

Denote the product of the entropy by $\tilde S$, i.e., $\widetilde{S}=S_{1}S_{2}...S_{n}=\prod\limits_{i=1}^{n} S_{i}$. Taking partial derivative of the entropy product with respect to the mass, we have
\be
\frac{\partial\widetilde{S}}{\partial M}=\widetilde{S}\left(\frac{1}{T_{1}S_{1}}+\frac{1}{T_2S_2}+... \right)=\widetilde{S} \sum\limits_{i=1}^{n}\frac{1}{T_{i}S_{i}}\,.
\ee

 Denote by $\hat S$ the entropy sum, i.e.,
\be
\hat{S}=S_{1}+S_{2}+...+S_{n}\,.
\ee
The first law yields
\be
\frac{\pa \hat{S}}{\pa M}=\sum\limits_{i=1}^{n}\frac{1}{T_{i}}\,.
\ee
Then it is straightforward to obtain the following theorem.
\begin{The}
For a black hole with multiple horizons, the entropy product is independent of the mass if and only if
\be
\sum\limits_{i=1}^{n}\frac{1}{T_iS_{i}}=0\,,
\ee
and the entropy sum is independent of the mass if and only if
\be
\sum\limits_{i=1}^{n}\frac{1}{T_{i}}=0\,.
\ee  \label{theo-first}
\end{The}

We shall see that this theorem is very useful in spherically symmetric spacetime because it does not require the specific form of metric.

\section{Criterion for spherical black holes in arbitrary dimensions}

The following discussion concerns spherical black holes described by the metric in the form
%**************fmetric
\be
ds^{2}=-f(r)dt^{2}+\frac{dr^{2}}{f(r)}+r^{2}d\Omega^{2} \label{fmetric}\,.
\ee
The roots of $f(r)$, denoted by $r_{1},r_{2},...,r_{n}$, are the radii of the horizons. Here the roots include complex ones, corresponding to the so called ``virtual horizons.''

%In a particular case that one root is zero, this equation still applies, however, if two or more roots are zero, this equation doesn't make sense since some denominators will be zero.

%In a particular case that one root is zero. Setting $r_{1}=0$, we have
%\be
%\sum\limits_{i=1}^{n} \prod\limits_{j\neq i}^{n}\frac{1}{r_{i}-r_{j}}=
%\ee

In general relativity, the temperature is proportional to $f'(r_i)$,
\be
T_{i}=\frac{1}{4\pi}f'(r_i)\,,
\ee
and the entropy is proportional to the area
\be
S_{i}=\frac{1}{4}A_{i}=\frac{1}{4}\Omega_{d-2}r^{d-2}\,,
\ee
where $\Omega_{d-2}$  is the area of a unit $(d-2)$ sphere
\be
\Omega_{d-2}=\frac{2\pi^{\frac{d-1}{2}}}{\Gamma(\frac{d-1}{2})}\,.
\ee

To proceed, we introduce a useful lemma first.
\begin{Lem} \label{lemma2}
Let $\{r_i\}$ be $n$ different numbers. Then
\be
\sum\limits_{i=1}^{n} \frac{r_{i}^{k}}{\prod\limits_{j\neq i}^{n}(r_{i}-r_{j})}=0\,,
\ee
where $0\leq k \leq n-2$.
\end{Lem}

For example, when $n=3$, the lemma gives
\be
\frac{1}{(r_1-r_2)(r_1-r_3)}+\frac{1}{(r_2-r_1)(r_2-r_3)}+\frac{1}{(r_3-r_1)(r_3-r_2)}=0\,,
\ee
and
\be
\frac{r_1}{(r_1-r_2)(r_1-r_3)}+\frac{r_2}{(r_2-r_1)(r_2-r_3)}+\frac{r_3}{(r_3-r_1)(r_3-r_2)}=0\,.
\ee
The proof of the lemma is a simple application of the Vandermonde determinant (see the Appendix for details).

Now let us assume that $f(r)$ is in a Laurent polynomial form,
\be
f(r)=1+a_{1}r+...+a_{n}r^{n}+b_{1}r^{-1}+...+b_{m}r^{-m}\,,
\ee
where $n,m\geq1$. Then we have
\be
r^{m}f(r)=r^{m}+a_{1}r^{m+1}+...+a_{n}r^{m+n}+b_{1}r^{m-1}+...+b_{m}=\prod\limits_{i}^{n+m}(r-r_{i})\,.
\ee
Taking derivative on both sides with respect to $r$ at each horizon, we have
\be\label{deri}
r_{i}^{m} f'(r_i)=\prod\limits_{j\neq i}^{n+m}(r_{i}-r_{j})\,.
\ee
This is equivalent to
\be
16\pi S_{i}T_{i}r_{i}^{m-(d-2)}=\Omega_{d-2}\prod\limits_{j\neq i}^{n+m}(r_{i}-r_{j})\,,
\ee
or
\be
\frac{1}{S_{i}T_{i}}\eqn \frac{16\pi r_{i}^{m-(d-2)}}{\Omega_{d-2}\prod\limits_{j\neq i}^{n+m}(r_{i}-r_{j})}\,,\\
\sum_i\frac{1}{S_{i}T_{i}}\eqn \sum_i\frac{16\pi}{\Omega_{d-2}} \frac{ r_{i}^{m-(d-2)}}{\prod\limits_{j\neq i}^{n+m}(r_{i}-r_{j})}\,. \label{siti}
\ee

According to Lemma \ref{lemma2}, as long as
$m-(d-2)\geq0$ and $m-(d-2)\leq m+n-2$, the right-hand side of \eq{siti} vanishes, and consequently, the entropy product is independent of the mass by virtue of Theorem \ref{theo-first}.

Similarly, \eq{deri} also gives
\be
\frac{1}{T_{i}}=\frac{4\pi r_{i}^{m}}{\prod\limits_{j\neq i}^{n+m}(r_{i}-r_{j})}\,,
\ee
which vanishes for $n\geq 2$.

Therefore, by applying Theorem \ref{theo-first} and Lemma \ref{lemma2}, we arrive at the following theorem:
\begin{The}\label{theorem-sp}
For a spherical black hole in a d-dimensional spacetime described by metric \meq{fmetric}, suppose f(r) is a Laurent polynomial:
\be
f(r)=1+a_{1}r+...+a_{n}r^{n}+b_{1}r^{-1}+...+b_{m}r^{-m}\,,
\ee
where $n$ and $m$ are positive integers.
If the entropy and temperature of each horizon located at $r=r_i$ are of the form
\be
S_i\eqn \frac{1}{4}{A_i}=\frac{1}{4}\Omega_{d-2}r^{d-2}\,,\\
T_i\eqn \frac{1}{4\pi}f'(r_i)\neq 0 \,;
\ee
then the necessary and sufficient condition for the entropy product being mass independent is that
%****************c1
\be
m\geq d-2 \ \ and\ \ \  n\geq 4-d. \label{c1}
\ee
The necessary and sufficient condition for the entropy sum being mass independent  is
%******************c2
\be
n\geq 2. \label{c2}
\ee
\end{The}

This theorem provides a very simple criterion to judge whether the entropy product/sum is mass independent. For example, $f(r) $  in a d-dimensional ($d\geq 4$) spherically symmetric Reissner-Nordstr\"om black hole takes the form
\be
f(r)=1-\frac{2M}{r^{d-3}}+\frac{Q^{2}}{r^{2(d-3)}}\,.
\ee
One sees immediately that
\be
m=2(d-3), \ \ \ n=0\,.
\ee
Therefore, $m\geq d-2$ and $n\geq 4-d$ are satisfied for $d\geq 4$. According to the theorem, the entropy product is independent of $M$ for all $d\geq 4$. Obviously, the condition $n\geq 2$ fails,  and consequently, the entropy sum must depend on $M$.

Now we consider charged black holes with cosmological constants. The solution is of the form
\be
f(r)=1-\frac{2M}{r^{d-3}}-\frac{2\Lambda}{(d-1)(d-2)}r^{2}+\frac{Q^{2}}{r^{2(d-3)}}\,.
\ee
When $Q=0$, it reduces to the Schwarzschild-de Sitter black hole, in which case
$m=d-3$, $n=2$. So $m\geq d-2$ fails, and the entropy product is mass dependent. This result has been found by Visser \cite{visser}.

For $Q\neq 0$, we have
\be
m=2(d-3), \ \ \  n=2\,.
\ee
We see that both \eqs{c1} and \meq{c2} are satisfied. Thus, the entropy product and entropy sum are both mass independent. This result agrees with Refs.\cite{cvetic, yu}.

\section{Myers-Perry black holes}

The entropy product/sum issue is not limited to spherical black holes.
 For Kerr-Anti-de Sitter black holes in four and higher dimensions\cite{gibbons}\cite{gibbons2}, Cveti\v{c} \emph{et al}.\cite{cvetic} showed that the entropy product of such black holes is independent of the mass  and Du and Tian\cite{yu} showed that the entropy sum is also mass independent.

Obviously, our Theorem \ref{theorem-sp} is not applicable to rotating black holes . However,  our Theorem \ref{theo-first} and Lemma \ref{lemma2} are still helpful for the mass dependence problem. In the following, we shall apply our technique developed above to  the  Myers-Perry solution \cite{myers}\cite{myers2}, which  is a generalization of the  Kerr solution in higher dimensions. It is necessary to discuss the Myers-Perry black holes in even and odd dimensions separately.

\subsection{Even dimensions}

Suppose $d=2n+2$ with $n\geq1$, in which case the metric for the Myers-Perry black hole is\cite{myers}
\be
ds^{2}=-dt^{2}+\frac{\mu r}{\Pi F}(dt+\sum\limits_{i=1}^{n}a_{i}\mu_{i}^{2}d\phi_{i})^{2}+\frac{\Pi F}{\Pi-\mu r}dr^{2}
+\sum\limits_{i=1}^{n}(r^{2}+a_{i}^{2})(d\mu_{i}^{2}+\mu_{i}^{2}d\phi_{i}^{2})+r^{2}d\alpha^{2}\,,
\ee
where
\be\label{F}
F(r)=1-\sum\limits_{i=1}^{n}\frac{a_{i}^{2}\mu_{i}^{2}}{r^{2}+a_{i}^{2}}\,,
\ee
%*************Pi
\be\label{Pi}
\Pi(r)=\prod\limits_{i=1}^{n}(r^{2}+a_{i}^{2})\,,
\ee
and $\alpha$ is an extra unpaired spatial coordinate.

There are $2n$ horizons located at the roots of the equation
%**************root
\be\label{root}
\Pi-\mu r=0\,.
\ee
Denote the $ith$ root by $r_i$. The corresponding horizon entropy is then given by
\cite{chen}
%*************si
\be
S_i=\frac{\Omega_{d-2}\Pi(r_{i})}{4}\,, \label{si}
\ee
and the surface gravity is given by\cite{myers}
%************kappai
\be
\kappa_i=\left.\frac{\pa_{r}\Pi-\mu}{2\mu r}\right|_{r=r_{i}} \,.\label{kappai}
\ee

To check the mass dependence of entropy product, we introduce the function
\be\label{starteven}
f(r)\equiv \left[\Pi(r)-\mu r\right]\frac{\Pi(r)}{2\mu r}\,.
\ee
Then it is not difficult to get
\be\label{fprime}
f'(r_i)=\left.\frac{\pa_{r}\Pi(r)-\mu}{2\mu r}\Pi(r)\right|_{r=r_{i}}=32\pi \Omega_{d-2}T_{i}S_{i}\,.
\ee

On the other hand, \eq{Pi} indicates that there are $2n$ roots for \eq{root}. Hence,
\eq{starteven} can be written in the form
\be
f(r)=\prod\limits_{j=1}^{2n}(r-r_{j})\cdot\frac{\Pi(r)}{2\mu r}\,,
\ee
and then
\be
f'(r_i)=\prod\limits_{j\neq i}(r_{i}-r_{j})\cdot\frac{\Pi(r_{i})}{2\mu r_{i}}=\oh\prod\limits_{j\neq i}(r_{i}-r_{j})\,,
\ee
where \eq{root} has been used in the last step. Together with \eqs{fprime}, we have
\be
\sum_i \frac{1}{T_iS_i}\sim \sum_{i=1}^{2n}\prod_{j\neq i}\frac{1}{(r_{i}-r_{j})}\,,
\ee
which vanishes according to Lemma \ref{lemma2}. Therefore,  the entropy product is independent of the mass of the black hole.

To discuss the entropy sum, we should start with the function
\be
\frac{\Pi-\mu r}{2\mu r}=\frac{\prod\limits_{i=1}^{2n}(r-r_{i})}{2\mu r}\,.
\ee

After taking derivatives with respect to $r$, we obtain
\be
\sum\limits_{i=1}^{2n}\frac{1}{8\pi T_{i}}=\sum\limits_{i=1}^{2n}\frac{2\mu r_{i}}{\prod\limits_{j\neq i}(r_{i}-r_{j})}\,,
\ee
which equals zero as long as $n\geq 2$, according to  Lemma \ref{lemma2}.
So the entropy sum is independent of the mass for $n\geq 2$. The only exception is $n=1$ ($d=4$) , i.e., the Kerr solution. In this case, it is straightforward to show that the entropy sum is $4M$.

\subsection{Odd dimensions}

Suppose $d=2n+1$, $d\geq 5$. The metric reads\cite{myers}
\be
ds^{2}=-dt^{2}+\frac{\mu r^{2}}{\Pi F}\left(dt+\sum\limits_{i=1}^{n}a_{i}\mu_{i}^{2}d\phi_{i}\right)^{2}+\frac{\Pi F}{\Pi-\mu r^{2}}dr^{2}
+\sum\limits_{i=1}^{n}(r^{2}+a_{i}^{2})(d\mu_{i}^{2}+\mu_{i}^{2}d\phi_{i}^{2})\,,
\ee
where $F$ and $\Pi$ are given by (\ref{F}) and (\ref{Pi}).
Again, there are $2n$ horizons located at the roots of
\be\label{rootodd}
\Pi-\mu r^{2}=0\,.
\ee
The surface gravity is\cite{myers}
\be
\kappa_i=\left.\frac{\pa_{r}\Pi-2\mu r}{2\mu r^{2}}\right|_{r=r_{i}}\,,
\ee
and the entropy is\cite{chen}
\be
S_i=\frac{\Omega_{d-2}\Pi(r_{i})}{4r_{i}}\,.
\ee

To discuss the entropy product, we start with the function
\be\label{startodd}
f(r)=[\Pi(r)-\mu r^{2}]\frac{\Pi(r)}{2\mu r^{3}}\,.
\ee
Then
\be
f'(r_i)=32\pi \Omega_{d-2}T_{i}S_{i}=\prod\limits_{j\neq i}(r_{i}-r_{j})\cdot\frac{\Pi(r_{i})}{2\mu r_{i}^{3}}\,.
\ee
Applying \eq{rootodd}, we can simplify the right-hand side as
\be
T_{i}S_{i}=\frac{\prod\limits_{j\neq i}(r_{i}-r_{j})}{2r_{i}}\,,
\ee
and then
\be
\frac{1}{32\pi \Omega_{d-2}}\sum\limits_{i=1}^{2n}\frac{1}{T_{i}S_{i}}=\sum_i\frac{2r_{i}}{\prod\limits_{j\neq i}(r_{i}-r_{j})}\,.
\ee
Since $d\geq 5$, $n\geq 2$, the entropy product is independent of the mass due to Theorem \ref{theo-first}.

To discuss the entropy sum, we should start with the term
\be
\frac{\Pi-\mu r}{2\mu r^{2}}=\frac{\prod\limits_{i=1}^{2n}(r-r_{i})}{2\mu r^{2}}\,.
\ee
By similar arguments, one can find
\be
\sum\limits_{i=1}^{2n}\frac{1}{8\pi T_{i}}=\sum_i^{2n}\left(\frac{2\mu r_{i}^{2}}{\prod\limits_{j\neq i}(r_{i}-r_{j})}\right)\,,
\ee
which equals zero as long as $n\geq 2$, according to Lemma 2.
So the entropy sum is independent of the mass for all odd dimensions ($d\geq 5$).

\section{Conclusions}

We have developed some criteria for the mass independence of black hole entropy product and entropy sum. By applying the first law of black hole thermodynamics, the explicit form of mass is no longer needed. Thus, our method allows the mass to take different forms as long as the first law is satisfied. This treatment is particularly useful for nonasymptotically flat spacetimes such as asymptotically (anti-) de Sitter spacetimes, where the mass can not be uniquely specified.

For  Reissner-Nordstr\"om black holes with or without cosmological constants,  the criterion becomes very simple and straightforward and does not require the detailed knowledge of the metric. The technique is also proven to be useful in nonspherically symmetric cases. For Myers-Perry black holes, we have shown that the entropy sum is mass independent for higher dimensions ($d>4$). It depends on the mass only in four dimensions, i.e., the Kerr solution.  With our method, it is also easy to demonstrate that the entropy product is mass independent for all dimensions.

\section*{Acknowledgements}
This research was supported by NSFC Grants No. 11235003, No.11375026 and No.NCET-12-0054.

\appendix

\section{Proof of the lemma}
Let $\{r_i\}$ be $n$ different numbers. We shall first show
%**************hzero
\be
h\equiv\sum\limits_{i=1}^{n} \prod\limits_{j\neq i}^{n}\frac{1}{r_{i}-r_{j}}=0 \,.\label{hzero}
\ee
By some algebra manipulations, $h$ can be written as
\be
h=\frac{\sum_ia_i}{\prod\limits_{1\leq i<j\leq n}(r_{i}-r_{j})}\,,
\ee
with
\be
a_i=(-1)^{i+1}\prod\limits_{j,k}(r_{j}-r_{k})\,,
\ee
where $1\leq j<k\leq n$ and $j,k\neq i$.

According to the Vandermonde determinant
\be
  \bv
  1&1&\dots&1\\
  r_{1}&r_{2}&\dots&r_{n}\\
  \hdotsfor{4}\\
  r_{1}^{n-1}&r_{2}^{n-1}&\dots&r_{n}^{n-1}\\
  \ev
  =\prod\limits_{n\geq i>j\geq 1}(r_{i}-r_{j})\,,
  %=(-1)^{C^{2}_{n}}\prod\limits_{1\leq i<j\leq n}(r_{i}-r_{j})
\ee
$a_i$ can be written as a $(n-1)\times(n-1)$ determinant,
\be
 a_i=K (-1)^{i+1}
  \bv
  1&\dots&1&1&\dots&1\\
  r_{1}&\dots&r_{i-1}&r_{i+1}&\dots&r_{n}\\
  \vdots&~&\vdots&\vdots&~&\vdots\\
  r_{1}^{n-2}&\dots&r_{i-1}^{n-2}&r_{i+1}^{n-2}&\dots&r_{n}^{n-2}\\
  \ev\,,
\ee
where
\be
K=(-1)^{C^{2}_{n-1}}\,,
\ee
is an overall constant independent of $i$.

So it is not difficult to verify that the sum of $a_i$ can be expressed as the  $n\times n$ determinant
\be
\sum_i a_i=K  \bv
  1&\dots&1&\dots&1\\
  1&\dots&1&\dots&1\\
  r_{1}&\dots&r_{i}&\dots&r_{n}\\
  \vdots&~&\vdots&~&\vdots\\
  r_{1}^{n-2}&\dots&r_{i}^{n-2}&\dots&r_{n}^{n-2}\\
  \ev \,,
\ee
which automatically vanishes. This completes the proof of \eq{hzero}.

Now we  prove the general formula
\be
h_k\equiv\sum\limits_{i=1}^{n} \prod\limits_{j\neq i}^{n}\frac{r_{i}^{k}}{r_{i}-r_{j}}=0\,,
\ee
where $k$ is any  integer satisfying $0\leq k\leq n-2$.

The first step is to rewrite $h_k$ as
\be
h_k=\frac{\sum_ib_i}{\prod\limits_{1\leq i<j\leq n}(r_{i}-r_{j})}\,,
\ee
with
\be
b_i=(-1)^{i+1}r^{k}_{i}\prod\limits_{j,k}(r_{j}-r_{k})\,.
\ee
By summing over $i$, we have a $n\times n$ determinant,
\be
\sum_i b_i=K
  \bv
  r_{1}^{k}&\dots&r_{i}^{k}&\dots&r_{n}^{k}\\
  1&\dots&1&\dots&1\\
  r_{1}&\dots&r_{i}&\dots&r_{n}\\
  \vdots&~&\vdots&~&\vdots\\
  r_{1}^{n-2}&\dots&r_{i}^{n-2}&\dots&r_{n}^{n-2}\\
  \ev\,,
\ee
which vanishes since $k$ is an integer and $0\leq k\leq n-2$. Therefore, we have completed the proof of Lemma \ref{lemma2}.

\end{document}